# Array Independent MIMO Channel Models with Analytical Characteristics

Yuan Yao, Jianfeng Zheng, and Zhenghe Feng, *Senior Member*, *IEEE*

## Abstract

The conventional analytical channel models for multiple-input multiple-output (MIMO) wireless radio channels are array dependent. In this paper, we present several array independent MIMO channel models that inherit the essence of analytical models. The key idea is to decompose the physical scattering channel into two parts using the manifold decomposition technique: one is the wavefield independent sampling matrices depending on the antenna arrays only; the other is the array independent physical channel that can be individually modeled in an analytical manner. Based on the framework, we firstly extend the conventional virtual channel representation (VCR), which is restricted to uniform linear arrays (ULAs) so far, to a general version applicable to arbitrary array configurations. Then, we present two *array independent* stochastic MIMO channel models based on the proposed new VCR as well as the Weichselberger model. These two models are good at angular power spectrum (APS) estimation and capacity prediction, respectively. Finally, the impact of array characteristics on channel capacity is separately investigated by studying the condition number of the array steering matrix at fixed angles, and the results agree well with existing conclusions. Numerical results are presented for model validation and comparison.

The authors are with the National Laboratory for Information Science and Technology, Department of Electric Engineering, Tsinghua University, Beijing, 100084 China (e-mails: {y-yao08, zjf98}@mails.tsinghua.edu.cn, fzh-dee@mail.tsinghua.edu.cn).

This work was partly supported by the National Science and Technology Major Project of the Ministry of Science and Technology of China under Contract 2009ZX03007003-02, and in part by ChuanXin Foundation of Tsinghua University.

## I. INTRODUCTION

Multiple-input multiple-output (MIMO) wireless technology has attracted considerable attentions in recent years due to its potential for a very high channel capacity [1]. Nevertheless, accurate and tractable channel modeling is essential for successful MIMO system design and deployment. A variety of channel models, mainly classified into *physical models* and *analytical models* in [2], have been proposed to understand and mimic the spatial characteristics of the MIMO channel. The physical models, e.g., [3]-[5], are quite accurate in reproducing the double-directional multipath propagation between the transmit array and receive array and are independent of the array configurations such as antenna pattern, number of elements, array geometry, mutual coupling, and polarization, *et al.* However, such models are usually of high complexity. On the other hand, the analytical models, e.g., [6]-[9], characterize the MIMO channel in an analytical way without explicitly accounting for the wave propagation, which are quite simple and easy-to-use. Nevertheless, these models are array dependent, that is, the results of the models are only applicable to the analysis of systems employing the same array configurations. Once the array conditions are changed, the model parameters have to be recalculated even for the same environment. Moreover, the impact of environment characteristics and array configurations on performances are inseparable in analytical models.

The virtual channel representation (VCR) [8], firstly proposed by Sayeed, is a relevant analytical model. It provides a tractable linear characterization of the physical scattering channel and offers a simple and geometric interpretation of the scattering environment. Such remarkable advantages make it very useful in many aspects such as space-time code design [10], [11], channel estimation [12], analysis of the capacity scaling behavior of MIMO channels [13], [14], and channel simulation, *et al*. However, this model is restricted to ideal uniform linear arrays (ULAs) so far, which hinders its application to real-world arrays with arbitrary structures and imperfections.

Several literatures have extended the conventional VCR in various aspects. Reference [13] presented a ULA-limited VCR for wideband MIMO channels. In [15], the uniform circular arrays (UCAs) and spherical arrays are also considered, but models for other array structures are still unavailable. The Weichselberger model [9], partially inspired by the conventional VCR, is applicable to arbitrary array configurations. However, the model is array dependent. The UIU model [16] and the recent canonical model [17] are similar to the Weichselberger model and they are also array dependent analytical models.

In this paper, we present several *array independent* MIMO channel models involving the essential characteristics of analytical models. By making use of the array manifold decomposition technique [18]-[21], the physical scattering channel is decomposed into two parts: one is the sampling matrices that have taken into account all the information of the arrays, and this part is independent of the environment characteristics; the other part is the array independent physical channel that can be separately modeled in an analytical way. Based on the decomposition, three specific contributions are presented.

Firstly, we propose an array independent VCR that is applicable to arbitrary array configurations. Unlike the conventional VCR, the spatial resolution of the new VCR does not depend on the number of array elements and their spacing but on the number of spatial basis functions only. The latter is mainly a tradeoff between the modeling accuracy and complexity. In addition, the new VCR gives a more direct reflection of the scattering environment since the virtual angles are just the azimuth angles without any transformation. Most importantly, many existing results [10]-[14], mainly based on the conventional ULA-restricted VCR, may be extended to general forms applicable to arbitrary array conditions by using the new VCR.

Then, two *array independent* stochastic MIMO channel models are presented. It has been shown in previous studies [9], [22], [23] that MIMO channel models may have different performance rankings under different metrics and the appropriate model has to be chosen according to the considered application. As for the our models, the first array independent stochastic model (AISM1), which is based on the new VCR, can provide accurate estimate of the 2-demensional (2-D) angular power spectrum (APS), but the accuracy of predicting the channel capacity is limited by the model complexity. On the other hand, the second model (AISM2), which is a straightforward extension of the Weichselberger model, can offer precise capacity prediction but less accurate APS estimation.

Finally, as the proposed models are array independent, the impact of array characteristics on channel performance can be separately investigated. For the first time, we use the condition number [24] of the array steering matrix at predefined angles to study the effect of array conditions on channel capacity. The ULAs as well as UCAs are considered as examples, and the results partly explain the existing conclusions.

This paper is organized as follows. In the next section, we present a general physical model for spatial MIMO channels, introduce the concept of manifold decomposition technique, and decompose the physical channel into wavefiled independent part and array independent part. Section III presents a new VCR for the array independent physical channel. The improvements compared to the conventional VCR are emphasized. In Section IV, two array independent stochastic MIMO channel models are presented based on the new VCR and the Weichselberger model. The extraction of the model parameters are discussed specifically. The impact of array characteristics on channel capacity is separately investigated In Section V. Finally, Section VI presents some numerical results for model validation and comparison. We conclude this paper in Section VII.

The notation used in this paper is given as follows.

- Boldface upper and lower case letters denote matrices and vectors, respectively.

- $\mathbb{C}^{N \times M}$ and $\mathbb{C}^{N \times 1}$ denote the sets of $N \times M$ complex matrices and $N$-dimensional complex vectors, respectively.

- Matrix $\mathbf{I}_N$ denotes an $N \times N$ identity matrix.

- Superscripts $(\cdot)^*$, $(\cdot)^T$, $(\cdot)^H$, $(\cdot)^{-1}$, and $(\cdot)^{-H}$ denote conjugate, transpose, conjugate transpose (Hermitian), inverse, and Hermitian inverse, respectively.

- $\otimes$ and $\odot$ denote Knoecker product and the element-wise product of two matrices (Hadamad product), respectively.

- $\det(\cdot)$, $\text{vec}(\cdot)$, $\|\cdot\|_F$ denotes the determinant operation, the vectorization operation (stacking the columns of the matrix into a vector), and Frobenius norm, respectively.

- $\text{E}[\cdot]$ denotes the expectation operation.

- $\lfloor x \rfloor$ gives the largest integer equal to or smaller than $x$. $\lceil x \rceil$ gives the smallest integer equal to or lager than $x$.

- $\delta(\cdot)$ denotes the Dirac delta function.

- Azimuth angle is represented by $\phi \in [-\pi, \pi)$.

## II. Channel Model

Consider a MIMO system with $N_T$ transmit antennas and $N_R$ receive antennas. Assume the channel is stationary and frequency-flat, the input-output relation of the MIMO channel is given by

$$\mathbf{y} = \mathbf{H}\mathbf{x} + \mathbf{n}, \tag{1}$$

where $\boldsymbol{x} \in \mathbb{C}^{N_T \times 1}$ is the transmit signal, $\boldsymbol{y} \in \mathbb{C}^{N_R \times 1}$ is the receive signal, $\boldsymbol{n} \in \mathbb{C}^{N_R \times 1}$ is the complex white zero-mean Gaussian noise, and $\mathbf{H} \in \mathbb{C}^{N_R \times N_T}$ denotes the channel matrix coupling the transmit and receive array elements.

In this paper, we consider the same physical scattering channel model as in [8] for simplicity, but we do not limit the array configurations. The model assumption does not restrict the generality of the discussion and the results can be extended to models including other channel parameters such as elevation angle, dual polarization, *et al.* See Section IV-D.

### A. Manifold Decomposition Technique

The manifold decomposition technique was initially proposed in [18] and has attracted much concern recently in array processing and direction finding [19]-[21]. The technique is in fact a framework for decomposition of the array steering vector on orthonormal basis functions. In this paper, we utilize it for MIMO channel modeling purpose.

Consider a 2-D antenna array of arbitrary structures and far-field scattering conditions. Define the origin of the coordinate system to be at the centroid of the array. The steering vector for the array with $N$ antenna elements is defined as $\boldsymbol{b}(\phi) = \left[ b_1(\phi),\, b_2(\phi),\, \ldots,\, b_N(\phi) \right]^T \in \mathbb{C}^{N \times 1}$, which denotes the signal response due to a far-field point source in the direction $\phi$. By using the manifold decomposition technique, $\boldsymbol{b}(\phi)$ can be expressed as [18]-[21]

$$\boldsymbol{b}(\phi) = \boldsymbol{\Gamma}\, \boldsymbol{d}(\phi) + \boldsymbol{\varepsilon}, \tag{2}$$

where $\boldsymbol{\Gamma} \in \mathbb{C}^{N \times M}$ is the sampling matrix that depends on the antenna array only, $\boldsymbol{d}(\phi) \in \mathbb{C}^{M \times 1}$ is the array independent spatial basis functions of the decomposition, $M \geqslant N$ is the number of the basis functions, $\boldsymbol{\varepsilon} \in \mathbb{C}^{N \times 1}$ denotes the modeling error. Note that $\boldsymbol{\varepsilon}$ can be safely neglected, provided that $M$ is a sufficiently large number [18]-[21].

There are several candidates for the spatial basis functions such as Fourier basis functions [19] and spherical harmonics [18], [20]. As indicated in [20], different basis functions are in fact equivalent to each other under some mild assumptions that typically hold in practice. Therefore, we use the Fourier decomposition for simplicity and $\boldsymbol{d}(\phi)$ is given by [19]

$$\boldsymbol{d}(\phi) = \frac{1}{\sqrt{M}} \left[ e^{j\frac{M-1}{2}\phi},\, \ldots,\, 1,\, \ldots,\, e^{-j\frac{M-1}{2}\phi} \right]^T. \tag{3}$$

Without loss of generality, we assume that $M$ is odd. Obviously, $\boldsymbol{d}(\phi)$ depends on the wavefield only and is independent of the array configurations.

The sampling matrix $\boldsymbol{\Gamma}$ contains all potential information of the array, including antenna pattern, element orientation and position, mutual coupling, and antenna manufacturing errors, *et al.*, and it is independent of the environment conditions. In practice, matrix $\boldsymbol{\Gamma}$ can be calculated from the discrete Fourier transform (DFT) of the array calibration measurement data (see [19] for details). However, we do not need to compute the sampling matrix in our models.

### B. Scattering Channel Model

For arbitrary array structures at the transmitter and receiver, the channel matrix $\mathbf{H}$ can be generally expressed as [8]

$$\mathbf{H} = \int_{-\pi}^{\pi} \int_{-\pi}^{\pi} G(\phi_R, \phi_T) \boldsymbol{b}_R(\phi_R) \boldsymbol{b}_T^H(\phi_T) d\phi_R d\phi_T , \tag{4}$$

where $G(\phi_R, \phi_T)$ denotes the *spatial spreading function*[1] that characterizes the scattering environment. $\boldsymbol{b}_T(\phi) \in \mathbb{C}^{N_T \times 1}$, $\boldsymbol{b}_R(\phi) \in \mathbb{C}^{N_R \times 1}$ are the steering vectors of the transmit and receive arrays, respectively.

The discrete version of the channel model in (4) is given by

$$\mathbf{H} = \sum_{l=1}^{L} \alpha_l \boldsymbol{b}_R(\phi_{R,l}) \boldsymbol{b}_T^H(\phi_{T,l}) , \tag{5}$$

where $\alpha_l$, $\phi_{T,l}$, and $\phi_{R,l}$ $(l = 1, 2, \ldots, L)$ denote the $l$-th independent complex path gains, the azimuth angles seen by the transmitter and receiver, respectively. $L$ denotes the total number of multipath components. For the discrete model, the spatial spreading function reduces to

$$G(\phi_R, \phi_T) = \sum_{l=1}^{L} \alpha_l \delta(\phi_R - \phi_{R,l}) \delta(\phi_T - \phi_{T,l}) . \tag{6}$$

By using the manifold decomposition technique, $\boldsymbol{b}_T(\phi)$ and $\boldsymbol{b}_R(\phi)$ can be expressed as

$$\boldsymbol{b}_T(\phi) = \boldsymbol{\Gamma}_T \boldsymbol{d}_T(\phi) + \boldsymbol{\varepsilon}_T ,$$

$$\boldsymbol{b}_R(\phi) = \boldsymbol{\Gamma}_R \boldsymbol{d}_R(\phi) + \boldsymbol{\varepsilon}_R , \tag{7}$$

---

[1] We do not distinguish the spatial domain and angular domain for brevity.

where $\boldsymbol{\Gamma}_T \in \mathbb{C}^{N_T \times M_T}$, $\boldsymbol{\Gamma}_R \in \mathbb{C}^{N_R \times M_R}$ are the sampling matrices of the transmit and receive arrays, respectively, $\boldsymbol{d}_T(\phi) \in \mathbb{C}^{M_T \times 1}$, $\boldsymbol{d}_R(\phi) \in \mathbb{C}^{M_R \times 1}$ (assume that $M_T$ and $M_R$ are odd) are the Fourier basis functions given by

$$\boldsymbol{d}_T(\phi) = \frac{1}{\sqrt{M_T}} \left[ e^{j\frac{M_T-1}{2}\phi}, \dots, 1, \dots, e^{-j\frac{M_T-1}{2}\phi} \right]^T,$$

$$\boldsymbol{d}_R(\phi) = \frac{1}{\sqrt{M_R}} \left[ e^{j\frac{M_R-1}{2}\phi}, \dots, 1, \dots, e^{-j\frac{M_R-1}{2}\phi} \right]^T, \tag{8}$$

$\boldsymbol{\varepsilon}_T \in \mathbb{C}^{N_T \times 1}$, $\boldsymbol{\varepsilon}_R \in \mathbb{C}^{N_R \times 1}$ denote the modeling errors of the transmit and receive arrays, respectively.

Since that the sampling matrices $\boldsymbol{\Gamma}_T$ and $\boldsymbol{\Gamma}_R$ are independent of the parameters of multipath components, the scattering channel model can be rewritten as

$$\mathbf{H} = \boldsymbol{\Gamma}_R \mathbf{H}_0 \boldsymbol{\Gamma}_T^H + \mathbf{H}_\varepsilon, \tag{9}$$

where the matrix $\mathbf{H}_0 \in \mathbb{C}^{M_R \times M_T}$ is given by

$$\mathbf{H}_0 = \int_{-\pi}^{\pi} \int_{-\pi}^{\pi} G(\phi_R, \phi_T) \boldsymbol{d}_R(\phi_R) \boldsymbol{d}_T^H(\phi_T) d\phi_R d\phi_T,$$

$$\mathbf{H}_0 = \sum_{l=1}^{L} \alpha_l \boldsymbol{d}_R(\phi_{R,l}) \boldsymbol{d}_T^H(\phi_{T,l}) \tag{10}$$

in the continuous and discrete forms, respectively. The matrix $\mathbf{H}_\varepsilon \in \mathbb{C}^{N_R \times N_T}$ denotes the error caused by the modeling errors $\boldsymbol{\varepsilon}_T$ and $\boldsymbol{\varepsilon}_R$, and is given by

$$\mathbf{H}_\varepsilon = \int_{-\pi}^{\pi} \int_{-\pi}^{\pi} G(\phi_R, \phi_T) \left( \boldsymbol{\Gamma}_R \boldsymbol{d}_R(\phi_R) \boldsymbol{\varepsilon}_T^H + \boldsymbol{\varepsilon}_R \boldsymbol{d}_T^H(\phi_T) \boldsymbol{\Gamma}_T^H + \boldsymbol{\varepsilon}_R \boldsymbol{\varepsilon}_T^H \right) d\phi_R d\phi_T$$

$$= \sum_{l=1}^{L} \alpha_l \left( \boldsymbol{\Gamma}_R \boldsymbol{d}_R(\phi_{R,l}) \boldsymbol{\varepsilon}_T^H + \boldsymbol{\varepsilon}_R \boldsymbol{d}_T^H(\phi_{T,l}) \boldsymbol{\Gamma}_T^H + \boldsymbol{\varepsilon}_R \boldsymbol{\varepsilon}_T^H \right) \tag{11}$$

where the second equality corresponds to the discrete model. Since $\boldsymbol{\varepsilon}_T$ and $\boldsymbol{\varepsilon}_R$ are ignorable for sufficiently large $M_T$ and $M_R$, we neglect $\mathbf{H}_\varepsilon$ for brevity and express the channel matrix $\mathbf{H}$ as

$$\mathbf{H} \approx \boldsymbol{\Gamma}_R \mathbf{H}_0 \boldsymbol{\Gamma}_T^H \tag{12}$$

Eq. (12) is the basis for all subsequent work in this paper. It indicates that the physical scattering channel can be decomposed into two separate parts: one ($\boldsymbol{\Gamma}_T$, $\boldsymbol{\Gamma}_R$) depends on the array characteristics only and is independent of the environment conditions; the other is the *array independent physical channel* ($\mathbf{H}_0$) which can be individually analyzed.

The matrix $\mathbf{H}_0$ in (10) can be interpreted as a MIMO channel with $M_T$ *virtual* transmit antennas and $M_R$ *virtual* receive antennas. The steering vectors of the virtual transmit and receive arrays are $\boldsymbol{d}_T(\phi)$ and $\boldsymbol{d}_R(\phi)$, respectively.

## III. Virtual Representation for the Array Independent Physical Channel

In this section, we present a new virtual representation for the array independent physical channel $\mathbf{H}_0$ (and hence the final physical channel $\mathbf{H}$). The main concept follows the idea of the conventional VCR in [8] but with some relevant differences, as shown in Table I. These differences are essentially due to different choices of the spatial basis functions, and they make the new VCR improved than the conventional one in many aspects:

- The conventional VCR is restricted to ULAs only, while the new VCR is array independent and hence is applicable to arbitrary array configurations.

- The spatial resolution of the conventional VCR is limited by the array apertures, which can make the approximation rather poor for a practical number of antenna elements. In contrast, the spatial resolution of the new VCR depends on the number of spatial basis functions only, which can be flexibly selected as a tradeoff between the model accuracy and complexity.

- The virtual angle defined in the conventional VCR is the *spatial angle* given by

$$\varphi = r\sin(\phi), \quad \phi = \sin^{-1}(\varphi/r) \tag{13}$$

where $r$ is the normalized (to wavelength) antenna spacing. The nonlinear transformation of the azimuth angle $\phi$ and its dependence on the antenna spacing make the spatial angle $\varphi$ a little inconvenient in relating the conventional VCR to the physical scattering channel immediately. On the other hand, the virtual angles in the new VCR are just the azimuth angles, and they are independent of the array configurations such as antenna spacing. Therefore, the new VCR reflects the characteristics of the scattering environment in a very direct manner.

In the following parts, we illustrate various aspects of the new VCR.

TABLE I

Main Differences between the New Virtual Channel Representation and the Conventional One

| Model | Array independent | Spatial resolution determined by | Virtual angle |
|---|---|---|---|
| Conventional VCR | No | Antenna spacing, $N_T$, $N_R$ | Spatial angle |
| New VCR | Yes | $M_T$, $M_R$ | Azimuth angle |

## A. Virtual Channel Representation

In (10), each propagation path is associated with an arbitrary transmit and receive azimuth angle. The virtual representation replaces the physical paths with virtual ones corresponding to *fixed virtual angles* that are determined by the spatial resolution of the spatial basis functions.

Define $\tilde{M}_T = (M_T - 1)/2$ and $\tilde{M}_R = (M_R - 1)/2$. By following the concept of the conventional VCR, the array independent physical channel $\mathbf{H}_0$ can be expressed as

$$\mathbf{H}_0 = \sum_{q=-\tilde{M}_R}^{\tilde{M}_R} \sum_{p=-\tilde{M}_T}^{\tilde{M}_T} \mathrm{H}_V(q,p) \boldsymbol{d}_R(\tilde{\phi}_{R,q}) \boldsymbol{d}_T^H(\tilde{\phi}_{T,p}) = \mathbf{D}_R \mathbf{H}_V \mathbf{D}_T^H, \tag{14}$$

where $\mathbf{H}_V \in \mathbb{C}^{M_R \times M_T}$ is the virtual channel representation, the matrices

$$\mathbf{D}_T = \left[ \boldsymbol{d}_T(\tilde{\phi}_{T,-\tilde{M}_T}), \, \ldots, \, \boldsymbol{d}_T(\tilde{\phi}_{T,\tilde{M}_T}) \right] \in \mathbb{C}^{M_T \times M_T},$$

$$\mathbf{D}_R = \left[ \boldsymbol{d}_R(\tilde{\phi}_{R,-\tilde{M}_R}), \, \ldots, \, \boldsymbol{d}_R(\tilde{\phi}_{R,\tilde{M}_R}) \right] \in \mathbb{C}^{M_R \times M_R} \tag{15}$$

are the DFT (unitary) matrices defined by fixed virtual angles that are uniformly distributed in azimuth domain

$$\tilde{\phi}_{T,p} = \frac{2\pi}{M_T} p, \quad -\tilde{M}_T \le p \le \tilde{M}_T,$$

$$\tilde{\phi}_{R,q} = \frac{2\pi}{M_R} q, \quad -\tilde{M}_R \le q \le \tilde{M}_R. \tag{16}$$

The spacing between the transmit/receive virtual angles denotes the spatial resolution of the new VCR

$$\Delta\phi_T = \frac{2\pi}{M_T}, \ \Delta\phi_R = \frac{2\pi}{M_R} \tag{17}$$

Fig. 1 illustrates the fixed virtual angles at the transmit side as well as $\Delta\phi_T$. Note that unlike the virtual angles used in the conventional VCR, the virtual angles in (16) are exactly fixed azimuth angles in the physical channel. Moreover, the virtual angles as well as the spatial resolution in (17) depend on the number of spatial basis functions only and are independent of the number of array elements and their spacing.

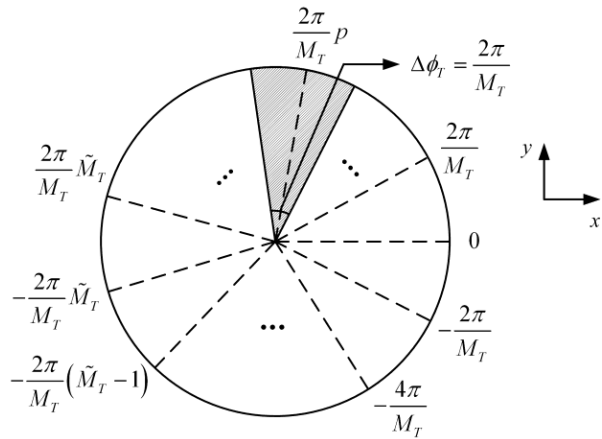

Fig. 1. Fixed virtual angles at the transmit side.

The virtual channel $\mathbf{H}_V = \mathbf{D}_R^H \mathbf{H}_0 \mathbf{D}_T$ is a unitary transformation of the array independent channel $\mathbf{H}_0$. The entry of $\mathbf{H}_V$ (Fourier series) is given by

$$\mathrm{H}_V(q,p) = \boldsymbol{d}_R^H\left(\tilde{\phi}_{R,q}\right)\mathbf{H}_0\boldsymbol{d}_T\left(\tilde{\phi}_{T,p}\right)$$

$$= \int_{-\pi}^{\pi}\int_{-\pi}^{\pi}G(\phi_R,\phi_T)f_{M_R}\left(\phi_R - \tilde{\phi}_{R,q}\right)f_{M_T}\left(\phi_T - \tilde{\phi}_{T,p}\right)d\phi_R d\phi_T$$

$$= \sum_{l=1}^{L}\alpha_l f_{M_R}\left(\phi_{R,l} - \tilde{\phi}_{R,q}\right)f_{M_T}\left(\phi_{T,l} - \tilde{\phi}_{T,p}\right), \tag{18}$$

where the last equality corresponds to the discrete model and

$$f_M(\phi) = \boldsymbol{d}^H(0)\boldsymbol{d}(\phi) = \frac{1}{M}\sum_{m=-(M-1)/2}^{(M-1)/2}e^{-jm\phi} = \frac{\sin(M\phi/2)}{M\sin(\phi/2)}. \tag{19}$$

To have a comparison with the conventional VCR, define [8]

$$g_N(\phi) = \frac{1}{N}\sum_{n=-(N-1)/2}^{(N-1)/2}e^{-j2\pi n\varphi} = \frac{\sin(\pi N\varphi)}{N\sin(\pi\varphi)}, \ \varphi = r\sin(\phi). \tag{20}$$

Plots of $\left|g_N(\phi)\right|$ and $\left|f_M(\phi)\right|$ versus the azimuth angle $\phi$ are shown in Fig. 2 (a) and (b), respectively. It can be seen that $\left|g_N(\phi)\right|$ has a strong dependence on the antenna spacing. For the conventional VCR, though the number of fixed virtual spatial angles in the principal period $\varphi \in [-0.5, 0.5]$ equals to $N$, the number of corresponding fixed azimuth angles in the range $\phi \in [-\pi/2, \pi/2]$, denoted as $N_a$, depends on the antenna spacing as follows:

- $N_a \lesssim N$ for $r < 0.5$ because no azimuth angle corresponds to the fixed virtual spatial angles outside the range $\varphi \in [-r, r] \subset [-0.5, 0.5]$. This may lead to a reduction in the spatial resolution of the conventional VCR.[2]

- $N_a = N$ for $r = 0.5$ since (13) is a one-to-one map between $\phi \in [-\pi/2, \pi/2]$ and $\varphi \in [-0.5, 0.5]$.

- $N_a \gtrsim N$ for $r > 0.5$ because the azimuth angles outside the range $\phi \in \left[ -\sin^{-1}(0.5/r), \sin^{-1}(0.5/r) \right]$ alias into the principal period $[-0.5, 0.5]$ of $\varphi$, which is called *spatial zooming and aliasing* [8] in the conventional VCR.

In all situations, $\left| g_N(\phi) \right|$ gets maximum values around the origin as well as $\pm\pi$ due to its periodicity. The zero points of $\left| g_N(\phi) \right|$ are nonuniformly distributed in the azimuth domain (dense around origin and sparse around $\pm\pi/2$), which results in nonuniform spatial resolution in the azimuth space.

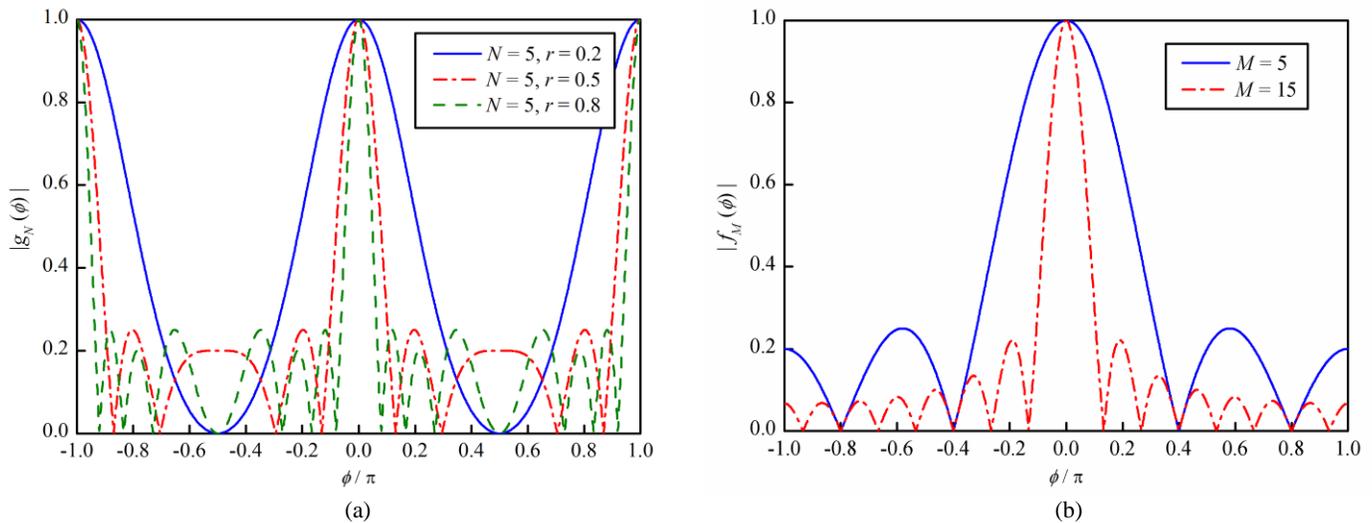

Fig. 2. Plots of (a) $\left| g_N(\phi) \right|$ and (b) $\left| f_M(\phi) \right|$ versus the azimuth angle $\phi$.

On the other hand, the results for the new VCR are much simpler. $\left| f_M(\phi) \right|$ gets peaky around the origin with increasing $M$, and it is zero at the other fixed virtual angles in (16) that are uniformly distributed in azimuth domain.

Thus, (18) indicates that the virtual channel coefficients are samples of a smoothed version of $G(\phi_R, \phi_T)$ at the fixed virtual azimuth angles. The smoothing is done by the kernel $f_{M_R}(\phi_R) f_{M_T}(\phi_T)$ that integrates to $4\pi^2 / M_T M_R$, and the smoothing kernel gets narrower around origin with increasing number of spatial functions.

---

[2] This situation is not discussed in [8].

*B. Virtual Path Portioning*

The virtual representation introduces a virtual path portioning that is very insightful in relating physical scattering characteristics to channel conditions. Define the following partition of the propagation paths

$$S_{T,p} = \left\{ l : -\frac{\pi}{M_T} \leq \phi_{T,l} - \frac{2\pi}{M_T} p < \frac{\pi}{M_T} \right\}, \quad -\tilde{M}_T \leq p \leq \tilde{M}_T,$$

$$S_{R,q} = \left\{ l : -\frac{\pi}{M_R} \leq \phi_{R,l} - \frac{2\pi}{M_R} q < \frac{\pi}{M_R} \right\}, \quad -\tilde{M}_R \leq q \leq \tilde{M}_R. \tag{21}$$

$S_{T,p}$ is the set of paths whose transmit azimuth angles are within $2\pi/M_T$ of the $p$-th virtual transmit angle $\tilde{\phi}_{T,p} = 2\pi p/M_T$ (the shadow area in Fig. 1). $S_{R,q}$ is similarly defined with respect to virtual receive azimuth angles. Note that

$$\bigcup_q S_{R,q} = \bigcup_p S_{T,p} = \bigcup_{p,q} \left( S_{T,p} \bigcap S_{R,q} \right) = \{1, 2, \ldots, L\}. \tag{22}$$

With the path portioning, the spatial spreading function for the discrete model can be expressed as

$$G(\phi_R, \phi_T) \approx \sum_{q=-\tilde{M}_R}^{\tilde{M}_R} \sum_{p=-\tilde{M}_T}^{\tilde{M}_T} \left[ \sum_{l \in S_{R,q} \bigcap S_{T,p}} \alpha_l \right] \delta(\phi_R - \tilde{\phi}_{R,q}) \delta(\phi_T - \tilde{\phi}_{T,p}), \tag{23}$$

and the virtual coefficients in (18) can be approximated as

$$H_V(q,p) \approx \left[ \sum_{l \in S_{R,q} \bigcap S_{T,p}} \alpha_l \right] \tag{24}$$

due to the property of $f_M(\phi)$, as shown in Fig. 2 (b). Eq. (23) and (24) indicate that the scattering contribution to the virtual azimuth angle pair $(\tilde{\phi}_{R,q}, \tilde{\phi}_{T,p}) = (2\pi q/M_R, 2\pi p/M_T)$ is proportional to the number of paths whose azimuth angles lie in the *virtual spatial bin* of size $(2\pi/M_R) \times (2\pi/M_T)$ centered on $(2\pi q/M_R, 2\pi p/M_T)$.

Thus, the multipath components are distributed in the virtual representation according to the spatial resolution.

*C. Virtual Representation of Realistic Channels*

The virtual presentation provides an intuitively appealing interpretation for realistic scattering environments, that is, different clusters correspond to different dominant sub-matrices of $\mathbf{H}_V$. Fig. 3 shows the contour plot of the normalized (to the maximum value) $E\left[ \left| H_V(q,p) \right| \right]$ for a simple synthetic scattering environment with cluster parameters listed in

Table II. $M_T = M_R = 101$, and $\mathbf{H}_0$ is generated via the discrete model in (10) using Gaussian path gains with zero mean. The expectation is over 200 independent channel realizations. It can be seen that the new VCR gives a direct and accurate imaging interpretation of the scattering environment. The accuracy of such geometry interpretation, i.e., the spatial resolution in (17), is governed by the number of spatial basis functions. For a given cluster with azimuth spreads $S_T = [\phi_{T-}, \phi_{T+}]$ and $S_R = [\phi_{R-}, \phi_{R+}]$, the size of the corresponding dominant sub-matrix of $\mathbf{H}_V$ is determined by

$$P_- \leq p \leq P_+ : \ P_- = \lfloor M_T \phi_{T-} \rfloor, \ P_+ = \lceil M_T \phi_{T+} \rceil,$$

$$Q_- \leq q \leq Q_+ : Q_- = \lfloor M_R \phi_{R-} \rfloor, \ Q_+ = \lceil M_R \phi_{R+} \rceil. \tag{25}$$

TABLE II

CLUSTER PARAMETERS OF A SYNTHETIC SCATTERING ENVIRONMENT

| Cluster index | Transmit/receive azimuth center /$\pi$ | Transmit/receive azimuth spread /$\pi$ | Number of paths within cluster |
|---|---|---|---|
| 1 | (-0.3, -0.3) | (0.1, 0.1) | 50 |
| 2 | (0, 0) | (0.1, 0.1) | 50 |
| 3 | (0.3, 0.3) | (0.1, 0.1) | 50 |

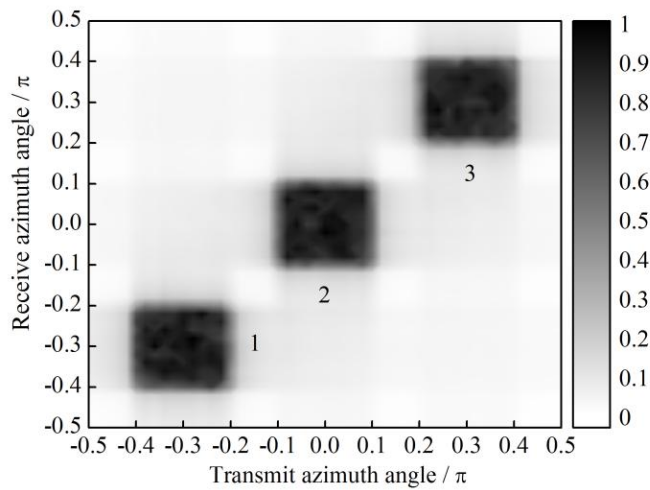

Fig. 3. Imaging interpretation of the scattering environment in Table II via $\mathbf{H}_V$. $M_T = M_R = 101$ and contour plot of $\mathrm{E}\left[\left|\mathrm{H}_V\left(q, p\right)\right|\right]$ is shown.

Note that clusters with different natures have quite different reflections in matrix $\mathbf{H}_V$ and hence different contributions to channel capacity and diversity. Interested readers are referred to [8], [9] for more details.

*D. Statistic of the Virtual Channel*

Discussions on statistics of **H** and $\mathbf{H}_V$ imposed by the physical scattering are very useful for our channel modeling work in Section IV. We make the same assumption on the physical scattering as in [8], i.e., the environment is *uncorrelated Rayleigh scattering*, that is, the spatial spreading function is an uncorrelated zero-mean Guassian random variable

$$\mathrm{E}\Big[G\big(\phi_R,\phi_T\big)G^*\big(\phi_R',\phi_T'\big)\Big]=M\big(\phi_R,\phi_T\big)\delta\big(\phi_R-\phi_R'\big)\delta\big(\phi_T-\phi_T'\big),\tag{26}$$

where $M\big(\phi_R,\phi_T\big)$ is the *spatial scattering function* that reflects the distribution of the channel power in $\big(\phi_R,\phi_T\big)$ domain. For the discrete physical model, the assumption corresponds to

$$\mathrm{E}\Big[\alpha_l\alpha_{l'}^*\Big]=\sigma_l^2\delta\big(l-l'\big).\tag{27}$$

Under the assumption, we obtain an important property of the virtual representation, that is, the entry of $\mathbf{H}_V$ is approximately uncorrelated (and hence statistic independent), which follows from (18)

$$\mathrm{E}\Big[\mathrm{H}_V\big(q,p\big)\mathrm{H}_V^*\big(q',p'\big)\Big]$$

$$=\int_{-\pi}^{\pi}\int_{-\pi}^{\pi}M\big(\phi_R,\phi_T\big)\times f_{M_R}\big(\phi_R-\tilde{\phi}_{R,q}\big)f_{M_R}\big(\phi_R-\tilde{\phi}_{R,q'}\big)\times f_{M_T}\big(\phi_T-\tilde{\phi}_{T,p}\big)f_{M_T}\big(\phi_T-\tilde{\phi}_{T,p'}\big)d\phi_R d\phi_T$$

$$\approx\frac{4\pi^2 M\big(\tilde{\phi}_{R,q},\tilde{\phi}_{T,p}\big)\delta\big(q-q'\big)\delta\big(p-p'\big)}{M_T M_R}$$

$$\approx\Bigg[\sum_{l\in S_{R,q}\cap S_{T,p}}\sigma_l^2\Bigg]\delta\big(q-q'\big)\delta\big(p-p'\big),\tag{28}$$

where the last approximation corresponds to the discrete model under virtual path partitioning. The approximation improves with increasing $M_T$ and $M_R$.

*E. Other Aspects*

In above parts, we have shown some aspects of the new VCR. Since the essential concept of the virtual representation is unchanged, we do not show the other aspects such as channel power distribution, channel correlation, $k$-diagonal virtual model in [8] or $D$-connected model in [13], *et al.*, for brevity. One can easily access these aspects of the new VCR following the analysis procedures in [8] but should also pay attention to the relevant differences between the new VCR and the conventional one, as listed in Table I. It is worth mentioning that the new VCR may extend many results of previous works, e.g., [10]-[14], to general forms applicable to arbitrary array configurations. In the following, we just show one of such potential extensions in channel modeling, i.e., new VCR channel models applicable to arbitrary array structures.

## IV. New Channel Models

In this section, we present two *array independent* stochastic models for MIMO channels. The first one (AISM1) is based on the new VCR described above, and the other one (AISM2) is based on the Weichselberger model considering joint correlation of both link ends.

### A. Array Independent Stochastic Model Based on the New Virtual Channel Representation (AISM1)

In Section III-D, we have proved that the entry of $\mathbf{H}_V$ is approximately uncorrelated (and hence statistic independent) under the assumption of uncorrelated Rayleigh scattering. Hence, let $\tilde{\mathbf{\Omega}}_{\text{angle}} = \left( \sigma_{qp} \right) \in \mathbb{C}^{M_R \times M_T}$, where

$$\sigma_{qp}^2 = \mathrm{E}\left[ \mathrm{H}_V\left(q,p\right) \mathrm{H}_V^*\left(q,p\right) \right] \approx \frac{4\pi^2 M\left(\tilde{\phi}_{R,q}, \tilde{\phi}_{T,p}\right)}{M_T M_R} \approx \left[ \sum_{l \in S_{R,q} \cap S_{T,p}} \sigma_l^2 \right] \tag{29}$$

denotes the variance of the entry of $\mathbf{H}_V$ according to (28). Eq. (29) indicates that the elements of $\tilde{\mathbf{\Omega}}_{\text{angle}}$ characterize the average power coupling between each pair of transmit and receive *virtual angles*. Hence, one can describe the channel spatial structure by specifying $\tilde{\mathbf{\Omega}}_{\text{angle}}$. From (14) and (29), $\tilde{\mathbf{\Omega}}_{\text{angle}}$ is also given by the element-wise square root of the following matrix

$$\mathbf{\Omega}_{\text{angle}} = \mathrm{E}\left[ \left( \mathbf{D}_R^H \mathbf{H}_0 \mathbf{D}_T \right) \odot \left( \mathbf{D}_R^T \mathbf{H}_0^* \mathbf{D}_T^* \right) \right]. \tag{30}$$

Then, $\mathbf{H}_V$ can be expressed in a stochastic form

$$\mathbf{H}_V = \tilde{\mathbf{\Omega}}_{\text{angle}} \odot \mathbf{G}, \tag{31}$$

where $\mathbf{G} \in \mathbb{C}^{M_R \times M_T}$ is a random matrix whose entries are i.i.d. complex Gaussian variables with zero-mean and unit variance. From (12) and (14), the channel matrix $\mathbf{H}$ can be expressed as

$$\mathbf{H} = \mathbf{\Gamma}_R \mathbf{D}_R \left( \tilde{\mathbf{\Omega}}_{\text{angle}} \odot \mathbf{G} \right) \mathbf{D}_T^H \mathbf{\Gamma}_T^H. \tag{32}$$

Define

$$\mathbf{B}_T = \mathbf{\Gamma}_T \mathbf{D}_T \in \mathbb{C}^{N_T \times M_T}, \quad \mathbf{B}_R = \mathbf{\Gamma}_R \mathbf{D}_R \in \mathbb{C}^{N_R \times M_R}. \tag{33}$$

From (7), (15), and ignore the modeling error $\boldsymbol{\varepsilon}$ again, $\mathbf{B}_T$ and $\mathbf{B}_R$ can be written as

$$\mathbf{B}_T = \left[ \boldsymbol{b}_T\left(\tilde{\phi}_{T,-\tilde{M}_T}\right), \ldots, \boldsymbol{b}_T\left(\tilde{\phi}_{T,\tilde{M}_T}\right)\right],$$

$$\mathbf{B}_R = \left[ \boldsymbol{b}_R\left(\tilde{\phi}_{R,-\tilde{M}_R}\right), \ldots, \boldsymbol{b}_R\left(\tilde{\phi}_{R,\tilde{M}_R}\right)\right]. \tag{34}$$

Thus, $\mathbf{B}_T$ and $\mathbf{B}_R$ are the transmit and receive *array steering matrices* corresponding to the predefined virtual angles in (16), respectively. Then, the channel matrix $\mathbf{H}$ can be expressed in a compact form (AISM1)

$$\mathbf{H}_{\text{AISM1}} = \mathbf{B}_R\left(\tilde{\boldsymbol{\Omega}}_{\text{angle}} \odot \mathbf{G}\right)\mathbf{B}_T^H. \tag{35}$$

### B. Array Independent Stochastic Model Based on the Weichselberger Model (AISM2)

Previous studies [9], [22], [23] have shown that the conventional VCR is good at estimating the 2-D APS but is less accurate for capacity prediction. On the contrary, the array dependent Weichselberger model shows excellent agreement with the measurement results in terms of capacity prediction, but it is less accurate for APS estimation. Here, we extend the Weichselberger model in a straightforward manner to an array independent version as a complementation model to the model AISM1.

Define the one-side correlation matrices of the array independent physical channel $\mathbf{H}_0$ as

$$\mathbf{R}_T = \mathrm{E}\left[\mathbf{H}_0^H\mathbf{H}_0\right] = \mathbf{U}_T\boldsymbol{\Lambda}_T\mathbf{U}_T^H \in \mathbb{C}^{M_T \times M_T},$$

$$\mathbf{R}_R = \mathrm{E}\left[\mathbf{H}_0\mathbf{H}_0^H\right] = \mathbf{U}_R\boldsymbol{\Lambda}_R\mathbf{U}_R^H \in \mathbb{C}^{M_R \times M_R}, \tag{36}$$

where $\mathbf{U}_T \in \mathbb{C}^{M_T \times M_T}$ and $\mathbf{U}_R \in \mathbb{C}^{M_R \times M_R}$ are unitary matrices whose columns are the eigenvectors of $\mathbf{R}_T$ and $\mathbf{R}_R$, respectively. $\boldsymbol{\Lambda}_T$ and $\boldsymbol{\Lambda}_R$ are diagonal matrices containing the eigenvalues of $\mathbf{R}_T$ and $\mathbf{R}_R$, respectively. Then, the array independent matrix $\mathbf{H}_0$ can be expressed as

$$\mathbf{H}_0 = \mathbf{U}_R\left(\tilde{\boldsymbol{\Omega}}_{\text{eigen}} \odot \mathbf{G}\right)\mathbf{U}_T^H. \tag{37}$$

The matrix $\tilde{\boldsymbol{\Omega}}_{\text{eigen}} \in \mathbb{C}^{M_R \times M_T}$ is the coupling matrix whose elements determine the average power coupling between each pair of transmit and receive *virtual eigenmodes* [9], and it is given by the element-wise square root of the following matrix

$$\boldsymbol{\Omega}_{\text{eigen}} = \mathrm{E}\left[\left(\mathbf{U}_R^H\mathbf{H}_0\mathbf{U}_T\right) \odot \left(\mathbf{U}_R^T\mathbf{H}_0^*\mathbf{U}_T^*\right)\right]. \tag{38}$$

From (12), the channel matrix $\mathbf{H}$ is given by

$$\mathbf{H} = \boldsymbol{\Gamma}_R \mathbf{U}_R \left( \tilde{\boldsymbol{\Omega}}_{\text{eigen}} \odot \mathbf{G} \right) \mathbf{U}_T^H \boldsymbol{\Gamma}_T^H . \tag{39}$$

Alternatively, from (33) $\mathbf{H}$ can be rewritten as (AISM2)

$$\mathbf{H}_{\text{AISM2}} = \mathbf{B}_R \mathbf{D}_R^H \mathbf{U}_R \left( \tilde{\boldsymbol{\Omega}}_{\text{eigen}} \odot \mathbf{G} \right) \mathbf{U}_T^H \mathbf{D}_T \mathbf{B}_T^H . \tag{40}$$

Note that unlike the Weichselberger model, the eigenmode coupling matrix $\tilde{\boldsymbol{\Omega}}_{\text{eigen}}$, the spatial eigenbasis $\mathbf{U}_T$ and $\mathbf{U}_R$ of the new model are all independent of the array configurations.

## C. Extraction of Model Parameters

All parameters needed for the MIMO channel models AISM1 and AISM2 are listed in Table III. In this part, we discuss some general principles in determining these model parameters.

TABLE III

PARAMETERS NEEDED FOR THE PROPOSED MIMO CHANNEL MODELS

| Model | Parameters |
|-------|-----------|
| AISM1 | $M_T, M_R, \mathbf{B}_T, \mathbf{B}_R, \ \hat{\boldsymbol{\Omega}}_{\text{angle}}$ |
| AISM2 | $M_T, M_R, \mathbf{B}_T, \mathbf{B}_R, \ \hat{\boldsymbol{\Omega}}_{\text{eigen}}, \mathbf{U}_T, \mathbf{U}_R$ |

Theoretically, the number of spatial basis functions (let $M_T = M_R = M$ for brevity) is a tradeoff between the model accuracy and complexity. The larger the number $M$ is, the more accurate the models represent for the real physical channel, and the more computation resources the models require. In practice, the computational accuracy of the used computer or the calibration measurement signal to noise ratio (SNR) may also be considered as factors that affect the $M$ selection because the modeling error $\boldsymbol{\varepsilon}$ in (2) is minimal for a finite $M$ rather than for $M \rightarrow +\infty$ in practical applications [19], [20]. Detailed error analysis of the models is out of the scope of this paper, and is considered as a future work.

The matrices $\mathbf{B}_T$ and $\mathbf{B}_R$ are the *array steering matrices* at the fixed virtual angles in (16). For arrays with simple geometry and ideal characteristics, e.g., ULAs and UCAs, the steering matrix can be conveniently calculated from the explicit expression of the array steering vector. For arrays with arbitrary structures and imperfections, one can obtain the steering matrix either by commercial simulation software or more precisely by measuring the array response to a far-field source located at the fixed virtual angles.[3] Note that for a given array, the steering matrix at fixed angles is a constant matrix that can be obtained offline.

The virtual angle coupling matrix $\tilde{\boldsymbol{\Omega}}_{\text{angle}}$ in AISM1, the eigenmode coupling matrix $\tilde{\boldsymbol{\Omega}}_{\text{eigen}}$, the eigenbasis $\mathbf{U}_T$ and $\mathbf{U}_R$ in AISM2 are all related to the array independent channel matrix $\mathbf{H}_0$ (see (30), (36), and (38)). In practice, the matrix $\mathbf{H}_0$ cannot be directly measured due to its inherently array independent characteristics. However, the following two alternative methods can be used to determine it.

*Method 1:* The first method is to calculate $\mathbf{H}_0$ from (10). The distributions of the physical channel parameters can be obtained either from existing physical MIMO channel models, e.g., [3]-[5], or from channel measurements using parameter estimation algorithms [25].

*Method 2:* The second method is to calculate $\mathbf{H}_0$ from the measurement data according to (12)

$$\mathbf{H}_0 = \boldsymbol{\Gamma}_R^{-1} \mathbf{H} \boldsymbol{\Gamma}_T^{-H}, \tag{41}$$

or equivalently

$$\mathbf{H}_0 = \mathbf{D}_R \mathbf{B}_R^{-1} \mathbf{H} \mathbf{B}_T^{-H} \mathbf{D}_T^H. \tag{42}$$

Some requirements should be met for (42) to hold. Firstly, the steering matrices $\mathbf{B}_T$ and $\mathbf{B}_R$ must be nonsingular square matrices, i.e., $M_T = N_T$, and $M_R = N_R$, which may result in relatively high modeling error $\boldsymbol{\varepsilon}$ and reduce the modeling accuracy. Secondly, we expect matrices $\mathbf{B}_T$ and $\mathbf{B}_R$ to be well-conditioned so that the numerical errors introduced by the matrix inverse operation are small [24]. Moreover, the number of the sounding array elements (also the number of spatial basis in this case) should be large enough to have a high spatial resolution in the measurement results. To satisfy these requirements, one should choose a special type of array whose steering matrix at predefined angles is well-conditioned when the number of array elements is relatively large and equals to the number of spatial basis functions. Fortunately, we have found that UCA is a good candidate for such channel sounding purpose (see Section V-B).

Once the matrix $\mathbf{H}_0$ is obtained, the model parameters $\tilde{\boldsymbol{\Omega}}_{\text{angle}}$, $\tilde{\boldsymbol{\Omega}}_{\text{eigen}}$, $\mathbf{U}_T$, and $\mathbf{U}_R$ can be calculated according to (30), (36), and (38), respectively. Note that once obtained, these model parameters need no recalculation any more for

---

a given environment since they are all array independent.

### D. Possible Extensions

The physical channel model (4) can be extended to more general cases. By following the analysis in [13], the model can be applied to wide band channels by adding fixed virtual delays. When taking elevation angle into account, the manifold decomposition in (2) needs to be modified as [20]

$$\boldsymbol{b}(\phi,\theta) = \boldsymbol{\Gamma}\boldsymbol{d}(\phi,\theta) + \boldsymbol{\varepsilon},$$  (43)

where $\theta$ denotes the elevation angle. Define

$$\boldsymbol{d}_{\phi}(\phi) = \frac{1}{\sqrt{M_{\phi}}}\left[e^{j\frac{M_{\phi}-1}{2}\phi}, \ldots, 1, \ldots, e^{-j\frac{M_{\phi}-1}{2}\phi}\right]^{T},$$

$$\boldsymbol{d}_{\theta}(\theta) = \frac{1}{\sqrt{M_{\theta}}}\left[e^{j\frac{M_{\theta}-1}{2}\theta}, \ldots, 1, \ldots, e^{-j\frac{M_{\theta}-1}{2}\theta}\right]^{T}.$$  (44)

Then, the spatial basis functions are given by

$$\boldsymbol{d}(\phi,\theta) = \boldsymbol{d}_{\phi}(\phi) \otimes \boldsymbol{d}_{\theta}(\theta).$$  (45)

The manifold decomposition considering dual polarization can be found in [21].

Moreover, the straightforward and easy extension of the Weichselberger model presented in Section IV-B implies that the other analytical models, e.g., [7], [16], [17], may be extended to array independent forms as well.

## V. The impact of Array Characteristics

The proposed models are very suitable for investigating the impact of array characteristics on channel performance since they are array independent. On one hand, for a given propagation environment, one can conveniently evaluate the performances under different array conditions by changing the array steering matrices at fixed angles only, i.e., matrices $\mathbf{B}_T$ and $\mathbf{B}_R$ in (34). On the other hand, the separation between the array configurations and the environment characteristics enables one to study the effect of array conditions individually. As an example, we use the condition number [24] of the array steering matrix at fixed angles as a measure to investigate the potential effect of array configurations on channel capacity. Note that the condition number of a given matrix was also used in [26] to analyze array robustness in sound field study.

For brevity, we consider the following general steering matrix at fixed angles instead of $\mathbf{B}_T$ and $\mathbf{B}_R$

$$\mathbf{B} = \left[ \boldsymbol{b}\left(\tilde{\phi}_{-\tilde{M}}\right), \ldots, \boldsymbol{b}\left(\tilde{\phi}_{\tilde{M}}\right) \right] \in \mathbb{C}^{N \times M} , \tag{46}$$

where $\tilde{M} = (M-1)/2$, and $\tilde{\phi}_n = 2\pi n/M$, $-\tilde{M} \leq n \leq \tilde{M}$. In theory, only matrix $\mathbf{B}$ for $M \to +\infty$ can reflect the true array characteristics without the modeling error $\boldsymbol{\varepsilon}$. However, since the error $\boldsymbol{\varepsilon}$ is ignorable for sufficiently large $M$, one can investigate matrix $\mathbf{B}$ with finite $M$ instead.

The condition number of matrix $\mathbf{B}$ is defined as [24]

$$\kappa(\mathbf{B}) = \|\mathbf{B}\|_F \cdot \|\mathbf{B}^{-1}\|_F = \frac{\beta_{\max}}{\beta_{\min}} , \tag{47}$$

where $\beta_{\max}$ and $\beta_{\min}$ denote the maximum and minimum (nonzero) singular values of matrix $\mathbf{B}$, respectively. Matrix $\mathbf{B}$ is called well-conditioned if $\kappa(\mathbf{B})$ is small (near 1), or ill-conditioned if $\kappa(\mathbf{B})$ is large.

As for our concern, for a stochastic scattering environment, an ill-conditioned matrix $\mathbf{B}$ is much likely to cause an ill-conditioned channel matrix $\mathbf{H}$ even if the array independent physical channel matrix $\mathbf{H}_0$ has a small condition number that corresponds to a rich scattering environment. As a result, the channel capacity possibly decreases because the singular values of matrix $\mathbf{H}$ are dispersed in a wide range [1], [6].

Since different types of arrays have quite different characteristics of the matrix $\mathbf{B}$, we select some examples to illustrate the concept.

### A. Uniform Linear Array

The steering vector of a $N$-element ULA with orientation angle $\phi_0$, as illustrated in Fig.4 (a), is given by

$$\boldsymbol{b}(\phi, \phi_0) = \left[ e^{j\pi(N-1)r\cos(\phi-\phi_0)}, \ldots, 1, \ldots, e^{-j\pi(N-1)r\cos(\phi-\phi_0)} \right]^T . \tag{48}$$

Obviously, $\kappa(\mathbf{B})$ is independent of $\phi_0$ for $M \to +\infty$. Thus, we assume $\phi_0 = \pi/2$ in the following.

As previous studies [1], [6], [8], [16] showed, the antenna spacing $r$ can have remarkable effect on channel capacity: a normalized antenna spacing no less than 0.5 is necessary for high capacity. Here, we provide a potential explanation for this conclusion via $\kappa(\mathbf{B})$.

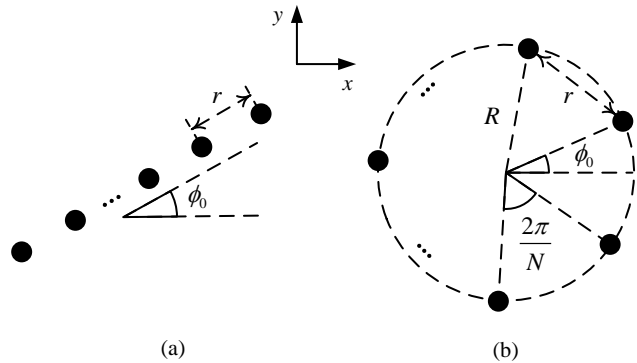

Fig. 4. Geometry and orientation of (a) a ULA, and (b) a UCA in $xy$-plane.

Consider a 5-element ULA with $\phi_0 = \pi/2$, Fig. 5 shows $\kappa(\mathbf{B})$ versus the normalized antenna spacing for different $M$ values. The large difference between $\kappa(\mathbf{B})$ for $M = 5$ and $M = 501$ is due to the effect of the modeling error $\boldsymbol{\varepsilon}$, and hence the results for $M = 5$ cannot be used to characterize the array characteristics. On the contrary, $\kappa(\mathbf{B})$ for $M = 11$ and $M = 501$ are very similar, which implies that the impact of the error $\boldsymbol{\varepsilon}$ is ignorable for $M \geqslant 11$ and $\kappa(\mathbf{B})$ for $M = 11$ is enough to represent the array conditions. In this case, $\kappa(\mathbf{B})$ is around 2 (well-conditioned) with very small variation for $r > 0.4$, but it increases almost linearly in the logarithmic scale as the spacing decreases for $r < 0.4$, i.e., matrix $\mathbf{B}$ is quite ill-conditioned for small antenna spacing. Such ill-conditioned matrix $\mathbf{B}$ is very likely to cause an ill-conditioned channel matrix $\mathbf{H}$ and reduce the channel capacity since the singular values of the spatial channels is dispersedly distributed, which partially explained why $r \geqslant 0.5$ is necessary for high capacity.

Moreover, the large error $\boldsymbol{\varepsilon}$ as well as $\kappa(\mathbf{B})$ for $M = N = 5$ indicate that ULA is not a good choice for *Method 2*.

### B. Uniform Circular Array

The steering vector of a $N$-element UCA, as illustrated in Fig.4 (b), is given by

$$\boldsymbol{b}(\phi) = \left[ e^{-j\pi r \frac{\cos(\phi + \pi(N-1)/N - \phi_0)}{\sin(\pi/N)}}, \ldots, e^{-j\pi r \frac{\cos(\phi - \pi(N-1)/N - \phi_0)}{\sin(\pi/N)}} \right]^T . \tag{49}$$

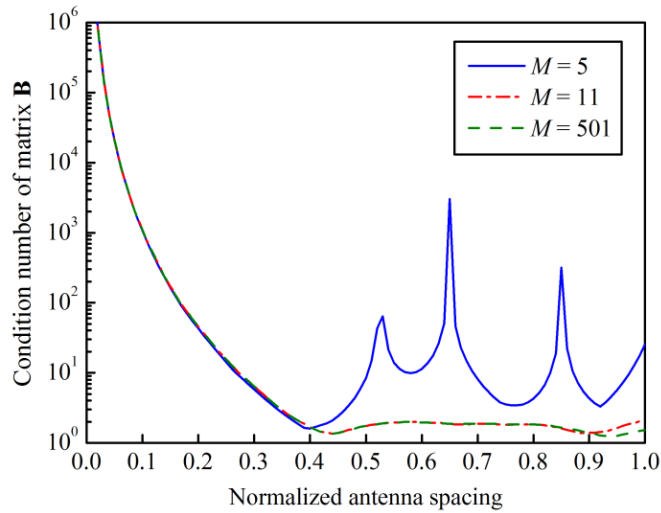

Fig. 5. Condition number of matrix **B** versus the normalized antenna spacing for a ULA with 5 antenna elements.

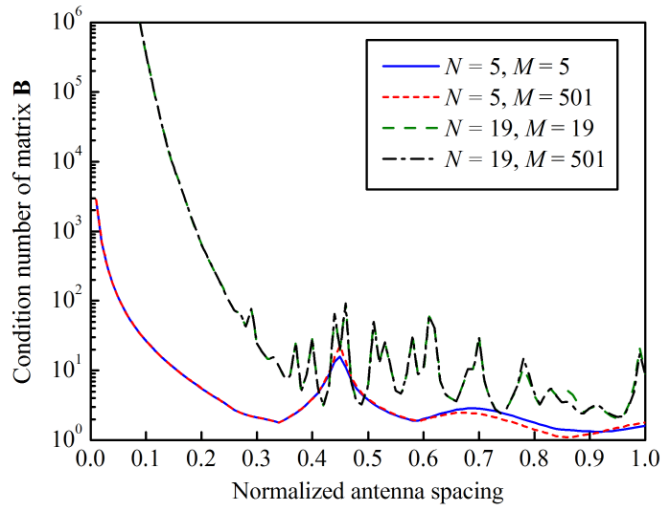

Fig. 6. Condition number of matrix **B** versus the normalized antenna spacing for UCAs with different numbers of array elements.

We choose $\phi_0 = 0$ for simplicity. Fig. 6 shows $\kappa(\mathbf{B})$ versus the normalized antenna spacing for UCAs with different numbers of array elements and spatial basis functions. It can be seen that $\kappa(\mathbf{B})$ for $M = N$ is almost the same as $\kappa(\mathbf{B})$ for sufficiently large $M$, which indicates that the modeling error $\boldsymbol{\varepsilon}$ has ignorable impact on $\kappa(\mathbf{B})$ and $\kappa(\mathbf{B})$ for $M = N$ is enough to characterize the characteristics of a UCA. Similar to the results in Fig. 5, $\kappa(\mathbf{B})$ decreases as the spacing increases, and $r > 0.4$ is desirable for small condition numbers. Furthermore, $\kappa(\mathbf{B})$ increase as the number of array elements increase. Note that $\kappa(\mathbf{B})$ is below 10 for $M = N = 19$ with $r = 0.5$, which indicates that UCA is a good candidate for *Method 2* since it satisfies the requirements of *Method 2* well. In Section VI, we show the feasibility of this method.

### C. Arbitrary Array Configurations

Above we have investigated two simple array types. Other regular arrays that have explicit steering vector expressions, such as triangle arrays, hexagon arrays, *et al.*, can be studied in a similar way. For much more practical antenna arrays with arbitrary configurations, the matrix **B** can be obtained either from simulation results or from measurement data and its condition number can also be considered as a potential performance measure in terms of capacity evaluations.

## VI. NUMERICAL RESULTS

In this section, we present some numerical results to validate the proposed models and show their attractive performances. For simplicity, we assume the transmit and receive arrays are of the same type, $\phi_0 = \pi/2$ for ULAs and $\phi_0 = 0$ for UCAs, and $N_T = N_R = N$, $M_T = M_R = M$ in the following.

### A. Validation of the Proposed Models

Firstly, the mutual information of MIMO channel is selected as a metric to validate the proposed models. Assume the transmitter has no knowledge about the channel at all, the channel capacity (in bits/s/Hz) is given by [1]

$$C = \log_2 \det\left(\mathbf{I}_N + \frac{\rho}{N}\mathbf{H}\mathbf{H}^H\right) \tag{50}$$

where $\rho$ is the evaluation SNR, the average energy of the channel matrix **H** is normalized such that $\|\mathbf{H}\|_F = N$ [1], [9]. $\rho$ is set to 20 dB for the following evaluations. The ergodic capacity is defend as $C_E = \mathrm{E}[C]$, where the expectation is over the statistics of the channel matrix **H**.

To have a comprehensive comparison, 100 different scenarios are generated. The number of clusters in each scenario is a random integer between 1 and 5. The mean transmit and receive azimuth angles $(\phi_T, \phi_R)$ and angle spreads $(\sigma_T, \sigma_R)$ of each cluster are uniformly distributed in the angle ranges $[-\pi, \pi)$ and $[0, \pi/2]$, respectively. 50 paths with zero-mean Gaussian gains are within each cluster, and the transmit (receive) azimuth angle associated with each multipath component is generated from the complex Gaussian process with mean $\phi_T$ ($\phi_R$) and variance $\sigma_T^2$ ($\sigma_R^2$). 200 independent channel realizations are used to calculate the ergodic capacity for each scenario.

5-element ULAs with antenna spacing $r = 0.5$ are used. In Fig. 7, we compare the conventional VCR (labeled as "Sayeed"), the Weichselberger model, and the proposed models AISM1 and AISM2 with the array independent physical channel $\mathbf{H}_0$ calculated by *Method 1* and *Method 2*. The number of spatial basis functions is chosen as $M = 19$. In *Method 2*, 19-element UCAs with antenna spacing $r = 0.5$ are used to sound the channel matrix $\mathbf{H}$, and $\mathbf{H}_0$ is obtained according to (42). The models AISM1 and AISM2 are then used to evaluate the performance of the MIMO systems with 5-element ULAs.

Fig. 7 shows the modeled ergodic capacity versus the true ergodic capacity via a scatter plot. Each point in the figure corresponds to a specific model and a specific scenario. The dash line indicates the points of no modeling error. It can be seen that the conventional VCR can accurately estimate the high ergodic capacity that corresponds to rich scattering scenarios, but it overestimates the low ergodic capacity seriously due to the low spatial resolution limited by a practical number of array elements ($N = 5$). The model AISM1 improves the estimation accuracy of the conventional VCR significantly due to its higher spatial resolution ($M = 19$), but it still tends to overestimate in the low capacity region. Note that the performance of model AISM1 can be further improved by increasing $M$, but the computational complexity increases as well. On the other hand, the model AISM2 has comparable estimation accuracy to the Weichselberger model, and they can precisely predict the true ergodic capacity of all scenarios. Moreover, the differences between the estimation results of the proposed models with *Method 1* and *Method 2* are very small, which indicates that UCA is a good choice for *Method 2* indeed. Thus, we only illustrate the performances of the proposed models with *Method 1* in subsequent tests for simplicity.

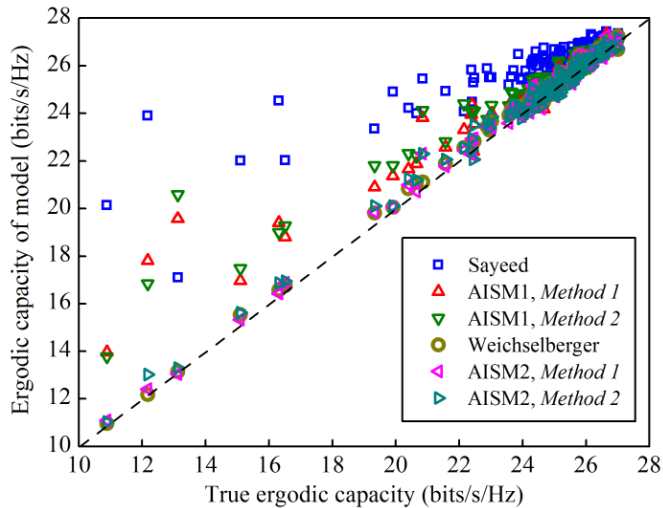

Fig. 7. Ergodic capacity for each of the 100 scenarios according to different MIMO channel models versus the true ergodic capacity. The identity line (dashed) indicates the points of no modeling error.

With the same array configurations as in Fig. 7, Fig. 8 shows the joint 2-D APS for the scenario described in Table II as well as the estimation results of different models. The 2-D APS is obtained by Capon's beamformer [25]

$$P_{\text{Capon}}\left(\phi_T, \phi_R\right) = \frac{1}{\tilde{\boldsymbol{b}}^H \mathbf{R}^{-1} \tilde{\boldsymbol{b}}} \tag{51}$$

where $\tilde{\boldsymbol{b}} = \boldsymbol{b}_T^*\left(\phi_T\right) \otimes \boldsymbol{b}_R\left(\phi_R\right)$, and $\mathbf{R} = \mathrm{E}\left[\mathrm{vec}\left(\mathbf{H}\right)\mathrm{vec}\left(\mathbf{H}\right)^H\right]$ denotes the full MIMO channel correlation matrix. 200 independent channel realizations are used. From Fig.8, one can seen that the model AISM1 shows a better fit to the true spectrum than the conventional VCR does due to higher spatial resolution of the former, whereas the Weichselberger model and the model AISM2 change the true APS much for this scenario.

### B. Impact of Array Characteristics

In this part, we show the convenience of using the proposed models to investigate the impact of array characteristics on channel performance. The cumulative distribution function (CDF) of capacity for the scenario described in Table II with different array configurations is considered. 1000 independent channel realizations are used to compute the CDF of capacity. Note that since the environment-related model parameters of the proposed models AISM1 and AISM2 have been obtained in last simulation, one does not need to compute them any more in the following tests. In contrast, the model parameters of the conventional VCR and the Weichselberger model have to be recalculated every time the array conditions are changed.

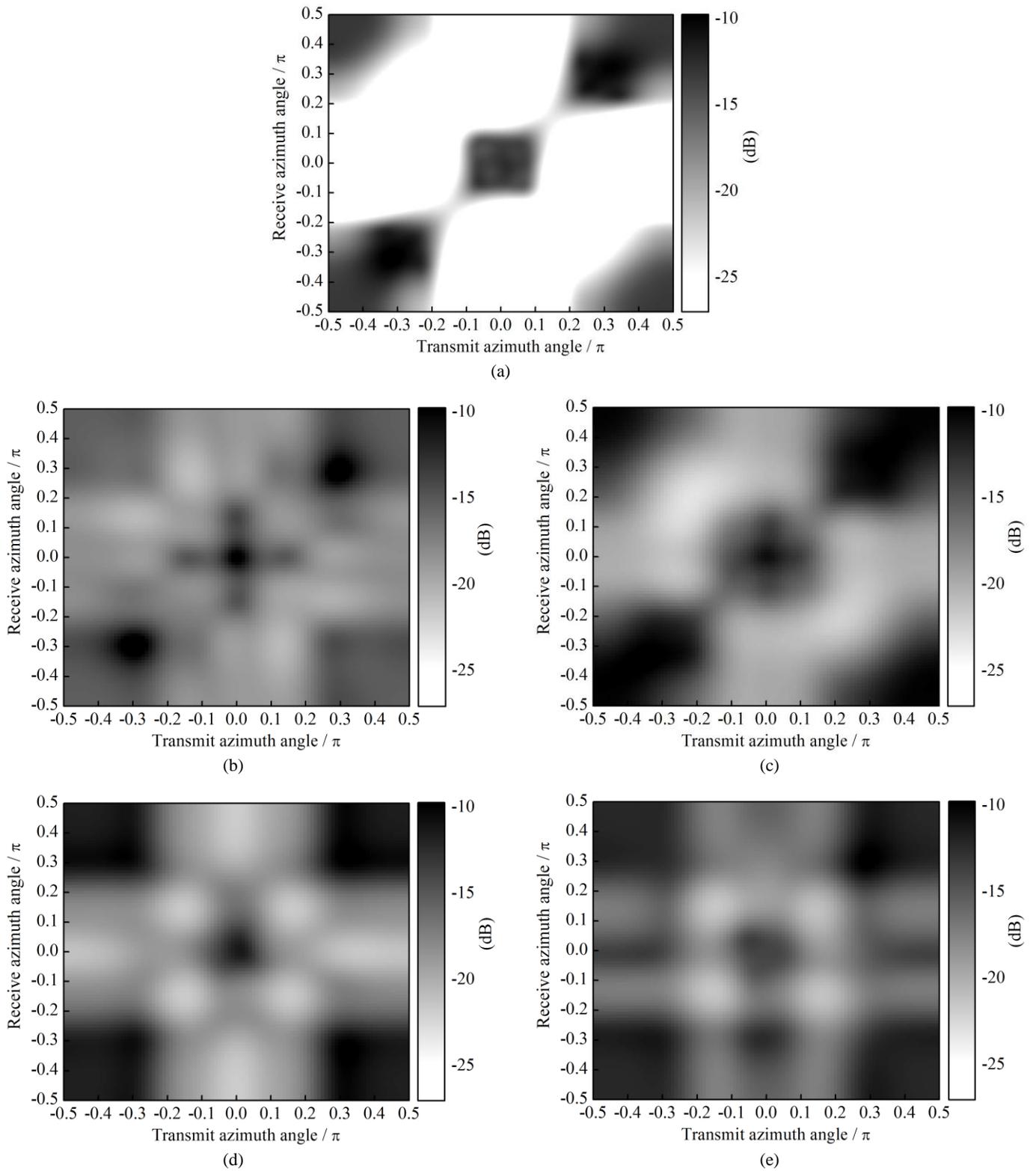

Fig. 8.  2-D angular power spectrum of (a) the scenario in Table II, (b) the conventional VCR, (c) the model AISM1 with *Method 1*, (d) the Weiselberger model, (e) the model AISM2 with *Method 1*.

Fig. 9 shows the CDF of capacity for 5-element ULAs with different antenna spacing,[4] $M = 19$ for models AISM1 and AISM2. Similar to the results of previous studies [8], larger antenna spacing results in larger capacity due to lower antenna correlation. The proposed models predict the capacity distribution accurately in both cases. The conventional VCR overestimates the capacity for $r = 0.2$ because its spatial resolution is very low in this case (see Fig.2 (a)).

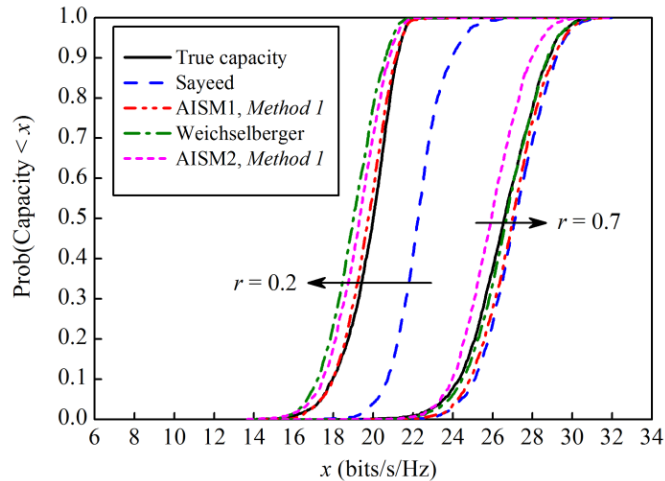

Fig. 9. CDF of capacity for the scenario in Table II. 5-element ULAs with different antenna spacing are used.

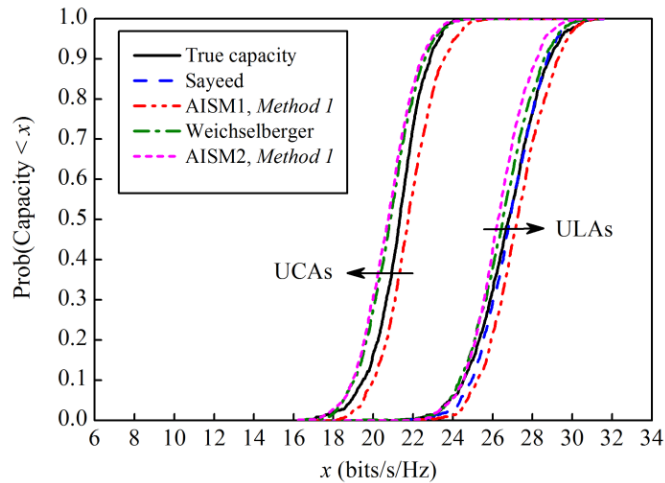

Fig. 10. CDF of capacity for the scenario in Table II. 5-element ULAs and UCAs with antenna spacing 0.5 are used.

Fig. 10 illustrates the CDF of capacity when different array types are used. 5-element ULAs and UCAs[5] with antenna spacing $r = 0.5$ are considered, $M = 19$ for models AISM1 and AISM2. All models predict the true CDF of capacity accurately. It can be seen that the MIMO system with ULAs at both link ends has higher capacity for this scenario.

---

[4] Note that more spatial basis functions are needed to model the array steering vector up to a desired accuracy when the antenna spacing increases because the array aperture increases as well [18]-[21].

[5] We do not apply the conventional VCR for the UCAs case.

$\kappa(\mathbf{B})$ for different array conditions in Fig. 9 and Fig. 10 as well as the mean condition number of the corresponding channel matrix are listed in Table IV. It can be seen that a larger $\kappa(\mathbf{B})$ corresponds to a larger $\mathrm{E}\left[\kappa(\mathbf{H})\right]$ and hence a lower ergodic capacity, which agrees well with previous analysis and partly explains the results in Fig. 9 and Fig. 10. Moreover, the results for the UCA case indicate that a low $\kappa(\mathbf{B})$ is likely to be a necessary but not sufficient condition for a well-conditioned channel matrix $\mathbf{H}$.

TABLE IV

CONDITION NUMBERS OF THE ARRAY STEERING MATRIX AND THE CORRESPONDING CHANNEL MATRIX

| Array configurations | $\kappa(\mathbf{B})$ | $\mathrm{E}\left[\kappa(\mathbf{H})\right]$ | $C_E$ (bits/s/Hz) |
|---|---|---|---|
| ULA, $r = 0.2$ | 46.04 | 5166.2 | 19.77 |
| ULA, $r = 0.7$ | 1.88 | 16.7 | 26.63 |
| ULA, $r = 0.5$ | 1.74 | 18.9 | 26.75 |
| UCA, $r = 0.5$ | 3.91 | 1259.1 | 21.22 |

## VII. CONCLUSIONS

Based on the manifold decomposition technique, this paper presents several array independent MIMO channel models with the essential characteristics of analytical models. The new VCR revisits the conventional one and improves the later in many aspects. Most importantly, the new VCR is applicable to arbitrary array configurations and its spatial resolution is unlimited by the array apertures, but is mainly a tradeoff between the modeling accuracy and complexity. The two stochastic MIMO channel models, which are based on the new VCR and the Weichselberger model, respectively, have extended the conventional models to array independent versions with comparable modeling accuracy. They are quite convenient for system evaluation and optimal array design. Finally, the impact of array characteristics on channel capacity is separately investigated via matrix analysis method, which gives alternative explanations for existing conclusions.

Further work needs to be done to refine the channel models in this paper. The possible extensions mentioned previously require actual realizations. Analyzing the effect of modeling error in (2) on the modeling accuracy is of importance. Besides the matrix analysis method, some other methods may be applied to investigate the impact of array characteristics on channel performance. Moreover, testing the ideas presented here in an experimental setup with real-world arbitrary arrays is quite desirable.